\documentclass[runningheads]{llncs}
\usepackage[T1]{fontenc}
\usepackage{graphicx}
\usepackage{amssymb}
\usepackage{amsmath}
\usepackage{color}
\usepackage{acronym}
\acrodef{LDPC}{low-density parity-check}
\acrodef{LDGM}{low-density generator matrix}
\acrodef{MDPC}{moderate-density parity-check}
\acrodef{QC}{quasi-cyclic}
\acrodef{QC-LDPC}{quasi-cyclic low-density parity-check}
\acrodef{QC-LDGM}{quasi-cyclic low-density generator matrix}
\acrodef{QC-MDPC}{quasi-cyclic moderate-density parity-check}
\acrodef{RSA}{Rivest-Shamir-Adleman}
\acrodef{BF}{bit flipping}
\acrodef{SPA}{sum product algorithm}
\acrodef{RDF}{random difference families}
\acrodef{ISD}{information set decoding}
\acrodef{KRA}{key recovery attack}
\acrodefplural{KRA}[KRAs]{key recovery attacks}
\acrodef{DA}{decoding attack}
\acrodefplural{DA}[DAs]{decoding attacks}
\acrodef{WF}{work factor}
\acrodef{BER}{bit error rate}
\acrodef{CER}{codeword error rate}
\acrodef{BSC}{binary symmetric channel}
\acrodef{BPSK}{binary phase shift keying}
\acrodef{$2$-PAM}{binary pulse amplitude modulation}
\acrodef{AWGN}{additive white Gaussian noise}
\acrodef{LLR}{log likelihood ratio}
\acrodefplural{LLR}[LLRs]{log likelihood ratios}
\acrodef{SPA}{sum-product algorithm}
\acrodef{DFR}{decoding failure rate}
\acrodef{SL}{security level}
\acrodef{ECC}{elliptic curve cryptography}
\acrodef{QD}{quasi-dyadic}
\acrodef{GRS}{generalized Reed-Solomon}
\acrodef{DSA}{Digital Signature Algorithm}
\acrodef{ECDSA}{Elliptic Curve Digital Signature Algorithm}
\acrodef{KEM}{key encapsulation module}
\acrodef{PKC}{public-key cryptosystem}
\acrodef{SK}{secret key}
\acrodef{PK}{public key}
\acrodef{IND-CCA}{indistinguishability under chosen ciphertext attack}
\acrodef{IND-CCA2}{indistinguishability under adaptive chosen ciphertext attack}
\acrodef{IND-CPA}{indistinguishability under chosen plaintext attack}
\acrodef{KI}{Kobara-Imai}
\acrodef{PFS}{perfect forward secrecy}
\acrodef{NP}{nondeterministic-polynomial}
\acrodef{DRBG}{deterministic random bit generator}
\acrodef{TRNG}{true random number generator}
\acrodef{KDF}{key derivation function}
\acrodef{AKE}{authenticated key exchange}
\acrodef{SDP}{syndrome decoding problem}

\newcommand{\sysacro}{SPANSE}

\newcommand{\Fq}{\mathbb F_q}

\usepackage{graphicx}
\usepackage{tikz}
\usetikzlibrary{calc, chains,
                fit,
                positioning,
                shapes}

\usetikzlibrary{automata,arrows,positioning,calc}
\definecolor{myblue}{RGB}{0,0,205}
\definecolor{mygreen}{RGB}{80,160,80}
\definecolor{myred}{RGB}{178,34,34}
\definecolor{mylightblue}{RGB}{0,153,153}
\definecolor{mygray}{RGB}{64,64,64}
\definecolor{myviolet}{RGB}{76,0,153}
\definecolor{myorange}{RGB}{204,102,0}

\usepackage{mathtools}

\begin{document}
\title{SPANSE: combining sparsity with density for efficient one-time code-based digital signatures}

\author{Marco Baldi \and
Franco Chiaraluce \and 
Paolo Santini}
\authorrunning{M. Baldi, F. Chiaraluce, P. Santini}
\institute{Università Politecnica delle Marche, Ancona, Italy\\\vspace{1em}
\email{\{m.baldi, f.chiaraluce, p.santini\}@univpm.it}
}
\maketitle

\begin{abstract}
The use of codes defined by sparse characteristic matrices, like QC-LDPC and QC-MDPC codes, has become an established solution to design secure and efficient code-based public-key encryption schemes, as also witnessed by the ongoing NIST post-quantum cryptography standardization process.
However, similar approaches have been less fortunate in the context of code-based digital signatures, since no secure and efficient signature scheme based on these codes is available to date.
The main limitation of previous attempts in this line of research has been the use of sparse signatures, which produces some leakage of information about the private key.
In this paper, we propose a new code-based digital signature scheme that overcomes such a problem by publishing signatures that are abnormally dense, rather than sparse.
This eliminates the possibility of deducing information from the sparsity of signatures, and follows a recent trend in code-based cryptography exploiting the hardness of the decoding problem for large-weight vectors, instead of its classical version based on small-weight vectors.
In this study we focus on one-time use and provide some preliminary instances of the new scheme, showing that it achieves very fast signature generation and verification with reasonably small public keys.

\end{abstract}
\section{Introduction}
The problem of decoding a random-looking linear code is considered as one of the most understood and well established problems in post-quantum cryptography.
Remarkably known to be NP-complete for both binary and non-binary codes \cite{Berlekamp1978,Barg1994}, the decoding problem is at the basis of several secure and efficient encryption schemes, stemming from the original proposals of McEliece \cite{McEliece1978} and Niederreiter \cite{Niederreiter1986}, or following more recent approaches \cite{Misoczki2013,AguilarMelchor2018}.
As an evidence, three encryption schemes (ClassicMcEliece, BIKE and HQC) out of the seven  currently admitted to the third round of the NIST post-quantum standardization process \cite{NISTcall2016} are based on codes.

Differently from such a rather consolidated scenario for encryption schemes, the situation is different concerning digital signature schemes.
There are basically two approaches to code-based digital signatures.
The first one, derived from the ``hash-and-sign'' paradigm used for instance in RSA signatures, encounters some obstacles when applied to the code-based setting.
This is due to the difficulty of randomly picking a decodable syndrome, yielding code-based schemes that are inefficient or insecure (or both).
Two historical proposals along this line are CFS \cite{Courtois01} and KKS \cite{Kabatianskii1997}, which however have important limitations.
In fact, it is very difficult to find secure though efficient instances of the KKS scheme \cite{Otmani2011}. 
The CFS scheme is more consolidated, but requires Goppa codes with unpractical parameters yielding very large public keys.
Moreover, Goppa codes with very high rate are required, which may expose the scheme to Goppa code distinguishers~\cite{Faugere2011}.
A recent important advance in this line of research is represented by the WAVE signature scheme~\cite{DebrisAlazard2019}, which however still requires a public key size in the order of 3 megabytes for 128 bits of classical security.
Interestingly, WAVE avoids previous attacks against signature schemes derived from CFS by replacing Goppa codes with some new codes with a special $(U, U+V)$ structure and relies on the hardness of decoding large-weight vectors, instead of the classical decoding problem looking for small-weight vectors. 

Other variants of CFS relying on codes different from Goppa codes have been proposed, but with less fortune.
In \cite{Baldi2013c}, a digital signature scheme based on \ac{QC-LDGM} codes was proposed, able to achieve very efficient signature generation and verification procedures, besides compact public keys.
Unfortunately, the use of sparse signatures in such a scheme turned out to produce some leakage of information concerning the private key, which lead to successful cryptanalysis \cite{Phesso2016}.

Opposed to the hash-and-sign paradigm used in the aforementioned signature schemes, a different approach to the design of code-based signatures is that of applying the Fiat-Shamir transform \cite{Fiat1987} to an identification scheme, which has the advantage of not relying on any trapdoor for key derivation.
In fact, consolidated zero-knowledge code-based identification schemes exist since a long time \cite{Stern1994}, which however exhibit significant soundness errors and thus require many repetitions to achieve reasonable security levels. This results in large signature sizes when they are used for digital signatures.
Subsequent variants of these schemes aim at overcoming such limitations \cite{Veron1997,Cayrel2011,Aguilar2011,ElYousfi2013,Bellini2019,barenghi2021less,Baldi2021,gueron2022designing,bettaieb2021zero}, although their features are still far from being comparable with those of signature schemes relying on other mathematical objects, such as lattices \cite{Lyubashevsky2012}.

\subsection{Our contribution}

We propose \sysacro{}, an evolution of the code-based signature scheme proposed in \cite{Baldi2013c}, following the hash-and-sign paradigm, with the aim of overcoming its main limitations and providing a new secure and efficient signature scheme based on a code-based trapdoor.
The main innovations over \cite{Baldi2013c} are as follows:
\begin{enumerate}
    \item Codes and vectors are defined over $\Fq$, with $q>2$ being a prime, while in \cite{Baldi2013c} they were binary.
    \item The transformation matrix $S$ was sparse in \cite{Baldi2013c}, while it is chosen to be dense in the new scheme, with entries from a subset of the underlying finite field.
    \item A transformation matrix $Q$ was needed in \cite{Baldi2013c} to hide the sparse nature of the other two secret matrices ($H$ and $S$), while it is no longer required in the new scheme, owing to the dense nature of $S$. For this reason, the matrix $Q$ is replaced with a simple permutation matrix $P$.
    \item The generated signatures were sparse in \cite{Baldi2013c}, while they are dense and free of zero entries in the new scheme.
\end{enumerate}

The above modifications allow avoiding attacks exploiting sparsity of signatures, which are no longer sparse, and make the scheme security rely on the hardness of decoding large-weight vectors instead of low-weight ones.
Still, some tuning of the overall statistical distribution of produced signatures will be needed in order to make it indistinguishable from a random distribution when long-term keys are used.
This, however, is left for future investigations, while in this paper we only focus on one-time keys.
The public matrices in \sysacro{} still have a product structure like those used in the LEDA cryptosystems \cite{Baldi2018,Baldi2019}. However, the dense nature of the multiplied matrices avoids the existence of weak keys like those identified in \cite{Apon2020}.
Codes in \ac{QC} form like in \cite{Baldi2013c} are used in the scheme we propose, and allow achieving public keys of reasonable size (in the order of 2 megabytes for 128 bits of classical security).
Such a public key size may be further reduced by increasing the circulant block size, which we keep very limited in this preliminary study.

One important advantage of the new scheme stemming from \cite{Baldi2013c} is that it exploits a straightforward decoding procedure, which makes signature generation, besides signature verification, very efficient from the computational standpoint.
This is a favourable aspect with respect to alternative code-based schemes resorting to classical decoding algorithms like WAVE, which uses a modified Prange decoder exploiting the $(U, U+V)$ code structure resulting in a signature generation complexity of order $O(\lambda^3)$, where $\lambda$ is the security level.
Notice that, in the one-time case we consider, the running time of our algorithm is dominated by the key generation procedure, which is still of order $O(\lambda^3)$, while signature generation has complexity reduced to order $O(\lambda^2)$.
However, even in the one-time case, key generation can be performed offline, so that its complexity is somewhat less important than the signature generation complexity.
Moreover, in the multiple-time case using long-term keys, the signature generation complexity is definitely more important than the key generation complexity.

As anticipated, in this work we focus on one-time use of the new scheme and show that it is secure against known attacks.
The study of multiple-time use, which will likely require the introduction of a final rejection sampling stage to avoid statistical attacks against long-term keys, is left as a future work.

\section{Preliminaries}
In this section we introduce the notation we will use throughout the paper and recall some basic notions concerning code-based cryptography.

\subsection{Linear codes}

Let $\Fq$ denote the Galois field of order $q$ and let $\Fq^{n}$ denote the $n$-dimensional vector space defined on $\Fq$.
A linear block code with length $n$ and dimension $k$, denoted as $\mathcal{C}\left(n,k\right)$, is a linear subspace of dimension $k$ of the vector space $\Fq^{n}$ formed by $n$-tuples defined over $\Fq$.
An encoding for the code $\mathcal{C}$ is a map $\Fq^{k}\mapsto \Fq^{n}$ that creates a unique association between any $k$-tuple (or information word) $u$ and an $n$-tuple (or codeword) $c$ belonging to the code $\mathcal{C}$.
Decoding instead starts from a version of $c$ corrupted by an error vector $e$, $\hat{c} = c + e$, and aims at recovering $c$ and $e$.

For any linear block code $\mathcal{C}\left(n,k\right)$, any set of $k$ linearly independent codewords $\left\{ g_{0}, g_{1},\ldots g_{k-1}\right\}$ forms a basis of the codeword space, such that any codeword
$c=\left[c_{0},c_{1},\ldots,c_{n-1}\right]$
can be expressed as a linear combination of the basis vectors:
\begin{equation}
c = u_{0} g_{0} + u_{1} g_{1} + \ldots + u_{k-1} g_{k-1}.
\label{eq:uiGi}
\end{equation}
The coefficients $u_{i}$ are taken from the information vector $u=\left[u_{0},u_{1},\ldots,u_{k-1}\right]$, such that encoding of $u$ into $c$ simply corresponds to the matrix operation $c = u G$, with
\begin{equation}
G = \left[\begin{array}{c} g_{0}\\ g_{1}\\ \vdots\\ g_{k-1} \end{array}\right]
\label{eq:uG}
\end{equation}
being a $k \times n$ matrix known as \textit{generator matrix} of the code $\mathcal{C}\left(n,k\right)$.

Any $k$ linearly independent codewords can be selected to form $G$, therefore any code can be provided with as many encodings as the overall number of possible generator matrices.
Encoding is said to be systematic if every codeword contains the information vector it is associated with.

A conventional form of systematic encoding is that in which each codeword is obtained by appending $r=n-k$ redundancy symbols to its corresponding $k$-symbol information word, that is
\begin{equation}
c = \left[u_{0},u_{1},\ldots,u_{k-1}|t_{0},t_{1},\ldots,t_{r-1}\right]
\label{eq:csystematic}
\end{equation}
It follows that the associated generator matrix $G$ can be written as
\begin{equation}
G = \left[I_k|P\right]\label{eq:Gsystematic}
\end{equation}
where $I_k$ denotes the $k\times k$ identity matrix and $P$ is a $k \times r$ general matrix.

Let us consider the orthogonal complement
$\Gamma^{\bot}$ of the set of codewords $\Gamma$. Its dimension is
\begin{equation}
r = \dim\left(\Gamma^{\bot}\right) = 
n - \dim\left(\Gamma\right) = n - k. 
\end{equation}
Given $r$ linearly independent $n$-symbol vectors $\left\{ h_{0},\ldots, h_{r-1}\right\}$
belonging to $\Gamma^{\bot}$, a basis of $\Gamma^{\bot}$ is simply obtained as
\begin{equation}
H = \left[\begin{array}{c}
h_{0}\\
h_{1}\\
\vdots\\
h_{r-1}
\end{array}\right]
\label{eq:H}
\end{equation}
and $\Gamma = \mathrm{Null}\left\{H\right\}$.
The matrix $H$ is known as a \textit{parity-check matrix} of the code $\mathcal{C}\left(n,k\right)$ and every codeword $c\in\Gamma$ must verify 
\begin{equation}
H c^T = 0_{r \times 1}
\label{eq:HcT}
\end{equation}
where $0_{r \times 1}$ represents the $r \times 1$ all-zero vector.
For any $n$-symbol vector $x \in \Fq^{n}$, the $r \times 1$ vector $s = H x^T$ is denoted as the \textit{syndrome} of $x$ through $H$.
It follows that a codeword belonging to $\mathcal{C}$ has an all-zero syndrome through $H$.

\subsection{QC-LDGM codes}

A linear block code is said to be \ac{LDGM} if at least one of its generator matrices is sparse, i.e., has a fraction of non-zero entries $\ll 1/2$. 
\sysacro{} uses a secret \ac{LDGM} code with length $n$ and dimension $k$, characterized by a generator matrix that is sparse and has non-zero entries taking small values (only $1$, as a special case).
The rows of $G$ have Hamming weight $w_g \ll n$.
Due to their sparse nature, it is very likely that, by summing two or more rows of the generator matrix of an LDGM code,
a vector with Hamming weight $> w_g$ is obtained.
If the linear combination of any group of rows of $G$ yields a codeword with weight greater than or equal to $w_g$, then the LDGM code has minimum distance $w_g$. 

The special class of \ac{LDGM} codes used in \sysacro{} is that of \ac{QC-LDGM} codes, having generator and parity-check matrices formed by circulant bocks with size $p \times p, p \in \left[2;r\right]$.
As already known, using matrices in this form allows reducing the memory needed to store them, and hence the public key size.
In fact, a \ac{QC} code is defined as a linear block code with dimension $k=p\cdot k_{0}$ and length $n=p\cdot n_{0}$, in which each cyclic shift of a codeword by $n_{0}$ symbols results in another valid codeword \cite{Townsend1967}.
From the definition of \ac{QC} codes it follows that the characteristic matrices of these codes can be written as formed by circulant blocks of the type:

\begin{equation}
A = \left[{\begin{array}{ccccc}
{a_{0}} & {a_{1}} & {a_{2}} & \cdots & {a_{p-1}}\\
{a_{p-1}} & {a_{0}} & {a_{1}} & \cdots & {a_{p-2}}\\
{a_{p-2}} & {a_{p-1}} & {a_{0}} & \cdots & {a_{p-3}}\\
\vdots & \vdots & \vdots & \ddots & \vdots\\
{a_{1}} & {a_{2}} & {a_{3}} & \cdots & {a_{0}}
\end{array}}\right].
\label{eq:CircMatrix}
\end{equation}

It is then evident that any circulant matrix can be fully described by just one of its rows or columns.
The set of $p \times p$ circulant matrices with entries from $\Fq$ forms a ring under the standard operations of matrix addition and multiplication over $\Fq$.
The zero element is the all-zero matrix, and the identity element is the $p \times p$ identity matrix.
If we consider the algebra of polynomials $\mathrm{mod}\left(x^{p}-1\right)$ over $\Fq$, $\Fq[x]/\langle x^p +1 \rangle$, the following map is an isomorphism between this algebra and that of $p \times p$ circulant matrices over $\Fq$:

\begin{equation}
A \leftrightarrow a\left(x\right)=\sum_{i=0}^{p-1}a_{i}\cdot x^{i}.
\label{eq:CircPoly}
\end{equation}
According to \eqref{eq:CircPoly}, any circulant matrix is associated to 
a polynomial in the variable $x$ having coefficients over $\Fq$ which coincide with the entries in the first row of the matrix: 
\begin{equation}
a\left(x\right)=a_{0}+a_{1}x+a_{2}x^{2}+a_{3}x^{3}+\cdots+a_{p-1}x^{p-1}.
\label{eq:CircPoly2}
\end{equation}
Similarly, the all-zero circulant matrix corresponds to the null polynomial and the identity
matrix to the unitary polynomial.

As it will be described next, the main part of the secret key of \sysacro{} is formed by a \ac{QC-LDGM} code described through its $k\times n$ generator matrix $G$ and $r\times n$ parity-check matrix $H$.
The values of $n$, $k$ and $r$ are all multiples of the circulant block size $p$.

For a \ac{QC-LDGM} code, the generator matrix $G$ takes the following general form

\begin{equation}
G = \left[
\begin{array}{ccccc}
G_{0,0} & G_{0,1} & G_{0,2} & \ldots & G_{0,n_0-1} \\
G_{1,0} & G_{1,1} & G_{1,2} & \ldots & G_{1,n_0-1} \\
G_{2,0} & G_{2,1} & G_{2,2} & \ldots & G_{2,n_0-1} \\
\vdots & \vdots & \vdots & \ddots & \vdots \\
G_{k_0-1,0} & G_{k_0-1,1} & G_{k_0-1,2} & \ldots & G_{k_0-1,n_0-1} \\
\end{array}
\right],
\label{eq:CQCLDGM}
\end{equation}
where each $G_{i,j}$ is a sparse circulant matrix or a null matrix
with size $p \times p$.
Hence, in this case the code length, dimension and redundancy are
$n=n_0p$, $k=k_0p$ and $r=(n_0-k_0)p=r_0p$, respectively.
Since a circulant matrix is defined by one of its rows (conventionally the first),
storing a matrix in the form \eqref{eq:CQCLDGM} requires $k_0 n_0 p$ symbols, yielding a reduction by a factor $p$ with respect to a matrix with a general form.

An important feature of the \ac{LDGM} codes used in \sysacro{} is that it is easy to obtain a random codeword $c$ belonging to the code and having weight approximately equal to a fixed, small value $w_c$.
Let us suppose, for simplicity, that $w_c$ is an integer multiple of $w_g$.
Since the rows of $G$ are sparse, it is very likely that, by summing a small number of rows, the Hamming weight of the resulting vector is about the sum of their Hamming weights.
Hence, by summing $\frac{w_c}{w_g}$ rows of $G$, chosen at random, we get a random codeword with Hamming weight about $w_c$.
It follows that the number of random codewords with weight close to $w_c$ 
can be roughly estimated as
\begin{equation}
A_{w_c} \approx {k \choose \frac{w_c}{w_g}}.
\end{equation}

\section{Dense code-based signatures
\label{sec:Scheme}}

\sysacro{} inherits the setting for the generation of sparse code-based signatures introduced in \cite{Baldi2013c}, with the main difference that the structure of the matrix $S$ is changed to obtain dense signatures intead of sparse ones.

As in \cite{Baldi2013c}, two public functions are fixed beforehand: a hash function $\mathcal{H}$ and a function $\mathcal{F}_\Theta$ that converts the output vector of $\mathcal{H}$ into a sparse binary vector $s$ with length $r$ and weight $w$ $(\le r)$. 
Note that, by sparse, we here refer to a vector taking values in $\{0 ; 1\}\subseteq \mathbb F_q$.
The vector $s$ is a public binary syndrome vector resulting from the signature generation procedure.
The output of $\mathcal{F}_\Theta$ is uniformly distributed and depends on a parameter $\Theta$, which is chosen for each message to be signed and is made public by the signer.
The choice of $\mathcal{F}_\Theta$ is discussed in Section \ref{sec:SignatureGen}.

\subsection{Key generation \label{sec:KeyGen}}

A first component of the secret key of \sysacro{} is a secret \ac{QC-LDGM} code having a generator matrix  of the type \eqref{eq:CQCLDGM}.
The overall row weight of the generator matrix $G$ of the secret code used in \sysacro{} is equal to $w_g$.
A \sysacro{} key pair is generated according to the following steps.

\begin{enumerate}
\item The signer randomly chooses a secret \ac{QC-LDGM} code with length $n = n_0p$ and dimension $k = k_0p$, having a generator matrix $G$ in the form \eqref{eq:CQCLDGM}, with row weight $w_g$, and from $G$ computes a systematic parity-check matrix $H$ for the same code.
Note that $G$ is binary as well, i.e., takes values in $\{0 ; 1\}\subseteq \mathbb F_q$.
\item Differently from \cite{Baldi2013c}, where the two secret matrices $Q$ and $S$ were used, in \sysacro{} the signer chooses two different secret matrices, $P$ and $S$, still in \ac{QC} form: an $r \times r$ secret permutation matrix $P$ and an $n \times n$ matrix $S$ with entries $\in \Fq$.
Having $S$ defined over $\Fq$ is another important difference of \sysacro{} with respect to \cite{Baldi2013c}.
In fact, in \cite{Baldi2013c} the matrix $S$ was sparse (besides \ac{QC}).
In \sysacro{}, instead, $S$ is still in \ac{QC} form but it is dense, with a different constraint of having small-valued entries taken from $\Fq$.
Let us denote as $d_i$ the fraction of entries equal to $i, i = 0,1,2,\ldots,q-1$, in each row of $S$. Then, we define
\begin{equation}
    d(x) = \sum_{i=0}^{q-1} d_i x^i,
\end{equation}
with $\sum_{i=0}^{q-1} d_i = 1$ as the polynomial describing the density of the symbols of $\Fq$ in each row of $S$.
As we will see next, the polynomials used in \sysacro{} as $d(x)$ are characterized by large values of $d_i$ for small values of $i$, and small or zero values of $d_i$ for large values of $i$. For brevity, we say that such polynomials are \textit{zero-concentrated}.
Note that this does not mean that $S$ is sparse. In fact, as we will see next, the overall density of non-zero entries in $S$  can easily be in the order of $50 \%$.
The matrix $Q$ used in \cite{Baldi2013c}, instead, is no longer needed in \sysacro{} because of the dense nature of $S$, and is replaced with a simple permutation matrix $P$.
\item The private key is $\left\{P,G, S\right\}$ and the public key is obtained as $H' = P^{-1} \cdot H \cdot S^{-1}$.
\end{enumerate}

\subsection{Signature generation \label{sec:SignatureGen}}

In \sysacro{}, the 
signature of a message ($m$) is computed as follows.
\begin{enumerate}
\item The signer computes $h = \mathcal{H}(m)$ and $s = \mathcal{F}_\Theta(h)$. The parameter $\Theta$ is chosen at random if statistical signatures are desired, otherwise it is fixed or computed as a one-way function of the message $m$ if deterministic signatures are desired.
The chosen value $\Theta^*$ is published along with the signature.
\item The signer computes the permuted syndrome $s' = Ps$. Then, being provided with the systematic parity-check matrix $H$ corresponding to the secret generator matrix $G$, the signer computes a sparse error vector $e$ having syndrome $s'$ through $H$.
Due to the systematic form of $H$, $e$ is easily obtained as $e = \left[ 0_{1 \times k} | s'^T \right]$, where $0_{1 \times k}$
is an all-zero row vector of length $k$.
\item The signer selects a random weight-$w_c$ codeword $c$ belonging to the secret \ac{QC-LDGM} code and computes the signature of $m$ as $\sigma = (e+c) \cdot S^T$.
If $\sigma$ has one or more zero entries, the signer chooses another random codeword $c$ and tries again, until $\sigma$ is free of zero entries.
\item The message $m$, the associated parameter $\Theta^*$ and the signature $\sigma$ are then published.
\end{enumerate}

An important parameter for any digital signature scheme is the total number of different signatures.
In \sysacro{}, a different signature corresponds to a different $r$-bit vector $s$, having weight $w$ and values in $\{0 ; 1\}$, so the total number of different signatures is
\begin{equation}
N_s = {r \choose w}.
\label{eq:signum}
\end{equation}

\subsection{Signature verification}

Signature verification in \sysacro{} is performed as follows.

\begin{enumerate}
\item The verifier checks that $\sigma$ is free of zero entries. If such a check fails, the signature is discarded.
\item Otherwise, the verifier computes $s^* = \mathcal{F}_{\Theta^*} (\mathcal{H}(m))$ and checks that $s^*$ has weight $w$.
If such a check fails, the signature is discarded.
\item Otherwise, the verifier computes $H' \cdot \sigma^T = P^{-1} \cdot H \cdot S^{-1} \cdot S \cdot (e^T + c^T) = P^{-1} \cdot H \cdot (e^T + c^T) = P^{-1} \cdot H \cdot e^T = P^{-1} \cdot s' = s$ and
compares it with $s^*$.
\item If $s = s^*$ the signature is accepted; otherwise, it is discarded.
\end{enumerate}

\section{Security analysis
\label{sec:Security}}
As anticipated, in this work we study the security of \sysacro{} under the simplifying assumption of one-time use, which is a necessary condition for a possible multiple-time use.
Since secure instances of \sysacro{} in the one-time case can easily be identified, as it will be shown next, the possible extension to multiple-time use will be the object of future works.

As a preliminary observation, we note that \sysacro{} lays its foundations on the scheme introduced in \cite{Baldi2013c}, with the crucial difference that attacks exploiting the sparsity of signatures are intrinsically avoided.
Among these, support decomposition attacks described in \cite{Baldi2013c} are the most dangerous ones, and are actually avoided in \sysacro{} by resorting to dense signatures.
The other attacks described in \cite{Baldi2013c} are still feasible, at least in principle, but their complexity remains exponential in the key size and can be easily rendered very large, as we will show next.

\subsection{Attacks based on decoding large-weight vectors}

Opposed to the system proposed in \cite{Baldi2013c}, which is based on the hardness of the classical decoding problem for low-weight vectors, the security of \sysacro{} relies on the difficulty of finding large-weight vectors through decoding of general codes.
In fact, general decoding algorithms could be used to:
\begin{itemize}
    \item mount key recovery attacks based on retrieving the rows of the dense generator matrix of the public code $G' = G S^T$, from which an attacker could try to recover the two secret matrices $G$ and $S$, or 
    \item mount forgery attacks by searching for a dense vector $\sigma_f$ corresponding to a given syndrome $s$ through the public parity-check matrix $H'$.  
\end{itemize}

In both these cases, an attacker should be able to find a vector free of zero entries\footnote{Note that, more generally, one may require vectors to have weight larger than some threshold value (which should be $\geq \frac{q-1}{q}n$).} that corresponds to a given syndrome (the all-zero vector for a key recovery attack and $s$ for a forgery attack) through the public parity-check matrix $H'$.
By using brute force, the probability of success for the attacker is easily computed as the fraction of valid vectors for each syndrome, that is,
\begin{equation}
    p_{BF} = \frac{(q-1)^n}{q^n q^r}.
\end{equation}
For example, by considering $n=15000$ and $q=127$, we have that $\left(\frac{q-1}{q}\right)^n < 2^{-170}$, which clearly highlights the unfeasibility of signature forgery attacks through brute force, even if the term $q^r$ is neglected.

So, as in WAVE, the security of our scheme relies on the hardness of decoding a given syndrome into a large-weight vector.
Differently from WAVE, however, we require signatures to have maximum weight.
A similar condition can also be imposed on the rows of the public code generator matrix $G' = G S^T$, as can be easily shown.
Obtaining vectors of full density is made easy in \sysacro{} by combining together sparsity and density in vector-matrix products.
Let us focus on signatures, but a similar reasoning also applies to the computation of $G' = G S^T$.
Generating signatures free of zero entries is made easy in \sysacro{} by the sparse nature of the vector $e+c$ combined with the dense and zero-concentrated nature of $S$.
To show this, let us consider $\Fq$ with $q$ being a prime and let us drop the modulo operation, considering for the moment that the operations are performed on the full set of integers.
The dense nature of $S$ makes the probability that any entry of the signature computed as $(e+c)S^T$ is exactly equal to zero negligible.
Furthermore, the sparse nature of $e+c$ together with the zero-concentrated nature of $S$ allow making the probability that any signature entry overcomes $q-1$ arbitrary small, or even null. 
Based on these considerations, when we restore the modulo operation we obtain that the probability of having a zero entry over $\Fq$ is negligible or even null.
This motivates the fact that generating signatures free of zero entries is easy for the signer.
Nevertheless, generating valid signatures may require some rejection sampling, depending on the choice of $d(x)$.
However, we show that when proper parameters are chosen, the number of needed attempts is very small.

\subsubsection{The PGE+SS framework}

In \cite{DebrisAlazard2019} the authors study the so-called PGE+SS framework to solve the large-weight vector decoding problem over the ternary finite field (i.e., for $q = 3$).
Their algorithm works by first applying Partial Gaussian Elimination (PGE) to reduce to a smaller instance, which is then solved through techniques inherited from the Subset Sum (SS) framework.
We extend the approach in \cite{DebrisAlazard2019} to the setting of interest for \sysacro{}, that is, considering vectors with entries over $\Fq$ with $q>3$ and full (or very large) density.
Starting from $s$ and $H$, a PGE can first be used to obtain
\begin{equation}H' = \begin{bmatrix}A\in\mathbb F_q^{(r-u)\times (n-u)} & 0_{(r-u)\times u}\in \mathbb F_q^{(r-u)\times u} \\
B\in\mathbb F_q^{u\times (n-u)} & I_{u}\in\mathbb F_q^{u\times u}\end{bmatrix},
\end{equation}
where $u = \varphi n$ with $0<\varphi<r = (1-R)n$.
We apply the same transformation to the syndrome $s$, obtaining $s' = [s_A, s_B]$, where $s_A$ has length $r-u$ and $s_B$ has length $u$. 
Let $e = \left[e_A, e_B\right]$ such that $He^T = s$. 
Then, we obtain
\begin{equation}
\label{eq:basic_relation}
\begin{cases}
Ae_A^T = s_A,\\
e_B^T = s_B - Be_A^T.
\end{cases}    
\end{equation}
To solve the above system, one first finds $e_A$ and then uses it to compute the corresponding $e_B$.
Since both $e_A$ and $e_B$ must have full weight, we have that $A$ and $s_A$ represent the input parity-check matrix and syndrome for another large weight decoding instance, with length $n' = n-u = (1-\varphi)n$ and redundancy $r' = r - u = (1-R-\varphi)n$.
We refer to it as the \textit{small instance} and use a SS solver to find solutions, i.e., candidates for $e_A$.

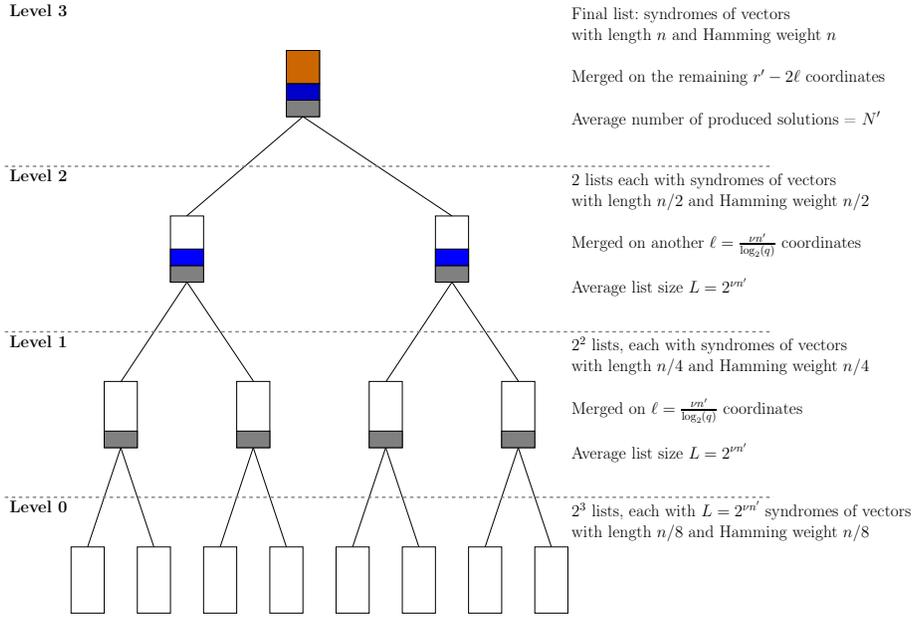
\begin{figure}
    \centering
    \newcommand{\blockdist}{2}
\newcommand{\blockheight}{2}
\newcommand{\blockwidth}{1}
\newcommand{\leveldist}{3}
\newcommand{\mergesize}{0.5}

\resizebox{\columnwidth}{!}{
\begin{tikzpicture}

\foreach \x in {0,...,3}{
\draw[thick] (-\x*\blockdist,0) rectangle (-\x*\blockdist+\blockwidth,\blockheight);
}

\foreach \x in {1,3}{
\draw[thick] (-\x*\blockdist+\blockwidth/2,\blockheight) edge (-\x*\blockdist+1.5*\blockwidth,\blockheight+\leveldist);
\draw[thick] (2*\blockwidth+3*\blockdist-\x*\blockdist+\blockwidth/2,\blockheight) edge (2*\blockwidth+3*\blockdist-\x*\blockdist+1.5*\blockwidth,\blockheight+\leveldist);
}

\foreach \x in {0,2}{
\draw[thick] (-\x*\blockdist+\blockwidth/2,\blockheight) edge (-\x*\blockdist-0.5*\blockwidth,\blockheight+\leveldist);
\draw[thick] (2*\blockwidth+3*\blockdist-\x*\blockdist+\blockwidth/2,\blockheight) edge (2*\blockwidth+3*\blockdist-\x*\blockdist-0.5*\blockwidth,\blockheight+\leveldist);
}

\foreach \x in {0,...,3}{
\draw[thick] (\blockdist+\x*\blockdist,0) rectangle (\blockdist+\x*\blockdist+\blockwidth,\blockheight);
}

\foreach \x in {0,...,1}{
\draw[thick] (-3*\blockdist+\blockwidth+2*\x*\blockdist,\leveldist+\blockheight) rectangle (-3*\blockdist+2*\blockwidth+2*\x*\blockdist,\leveldist+2*\blockheight);
\draw[fill = gray] (-3*\blockdist+\blockwidth+2*\x*\blockdist,\leveldist+\blockheight) rectangle (-3*\blockdist+2*\blockwidth+2*\x*\blockdist,\leveldist+\blockheight+\mergesize);
}

\foreach \x in {1,3}{
\draw[thick] (-3*\blockdist+1.5\blockwidth+2*\x*\blockdist,\leveldist+2*\blockheight) edge (-3*\blockdist+2.5*\blockwidth+2*\x*\blockdist-\blockwidth-\blockdist,2*\leveldist+2*\blockheight);
\draw[thick] (-3*\blockdist+1.5\blockwidth+2*\x*\blockdist-2*\blockwidth-\blockdist,\leveldist+2*\blockheight) edge (-3*\blockdist+2.5*\blockwidth+2*\x*\blockdist-\blockwidth-\blockdist,2*\leveldist+2*\blockheight);
}

\foreach \x in {0,...,1}{
\draw[thick] (2*\blockwidth + \blockwidth+2*\x*\blockdist,\leveldist+\blockheight) rectangle (2*\blockwidth +2*\blockwidth+2*\x*\blockdist,\leveldist+2*\blockheight);
\draw[fill = gray] (2*\blockwidth + \blockwidth+2*\x*\blockdist,\leveldist+\blockheight) rectangle (2*\blockwidth +2*\blockwidth+2*\x*\blockdist,\leveldist+\blockheight+\mergesize);
}

\draw[thick] (-\blockdist-\blockwidth,2*\leveldist+2*\blockheight) rectangle (-\blockdist,2*\leveldist+3*\blockheight);
\draw[fill = gray] (-\blockdist-\blockwidth,2*\leveldist+2*\blockheight) rectangle (-\blockdist,2*\leveldist+2*\blockheight+\mergesize);
\draw[fill = blue] (-\blockdist-\blockwidth,2*\leveldist+2*\blockheight+\mergesize) rectangle (-\blockdist,2*\leveldist+2*\blockheight+2*\mergesize);

\draw[thick] (-\blockdist-0.5\blockwidth,2*\leveldist+3*\blockheight) edge (\blockwidth,3*\leveldist+3*\blockheight);

\draw[thick] (\blockdist+3.5*\blockwidth, 2*\leveldist+3*\blockheight) edge (\blockwidth,3*\leveldist+3*\blockheight);

\draw[thick] (\blockdist+3*\blockwidth,2*\leveldist+2*\blockheight) rectangle (\blockdist+4*\blockwidth,2*\leveldist+3*\blockheight);
\draw[fill = gray] (\blockdist+3*\blockwidth,2*\leveldist+2*\blockheight) rectangle (\blockdist+4*\blockwidth,2*\leveldist+2*\blockheight+\mergesize);
\draw[fill = blue] (\blockdist+3*\blockwidth,2*\leveldist+2*\blockheight+\mergesize) rectangle (\blockdist+4*\blockwidth,2*\leveldist+2*\blockheight+2*\mergesize);

\draw[thick] (-\blockwidth/2+\blockdist/2,3*\leveldist+3*\blockheight) rectangle (\blockwidth/2+\blockdist/2,3*\leveldist+4*\blockheight);

\draw[fill = gray] (-\blockwidth/2+\blockdist/2,3*\leveldist+3*\blockheight) rectangle (\blockwidth/2+\blockdist/2,3*\leveldist+3*\blockheight+\mergesize);
\draw[fill = myblue] (-\blockwidth/2+\blockdist/2,3*\leveldist+3*\blockheight+\mergesize) rectangle (\blockwidth/2+\blockdist/2,3*\leveldist+3*\blockheight+2*\mergesize);
\draw[fill = myorange] (-\blockwidth/2+\blockdist/2,3*\leveldist+3*\blockheight+2*\mergesize) rectangle (\blockwidth/2+\blockdist/2,3*\leveldist+4*\blockheight);


\draw[very thick, dashed, gray] (-2-2*\blockwidth-2*\blockdist,\blockheight + \leveldist/2) edge (6.2+3*\blockwidth+3*\blockdist,\blockheight + \leveldist/2);

\node [anchor = north east ]at(-2*\blockwidth-2*\blockdist,\blockheight + \leveldist/2){$\Large \textbf{Level 0}$};

\node [anchor = north west ]at(3*\blockwidth+3*\blockdist,\blockheight + \leveldist/2){\Large $\begin{matrix*}[l]\text{$2^3$ lists, each with $L = 2^{\nu n'}$ syndromes of vectors}\\
\text{with length $n/8$ and Hamming weight $n/8$}\end{matrix*}$};

\draw[very thick, dashed, gray] (-2-2*\blockwidth-2*\blockdist,\blockheight + \leveldist/2 + \blockheight + \leveldist) edge (6.2+3*\blockwidth+3*\blockdist,\blockheight + \leveldist/2+ \blockheight + \leveldist);

\node [anchor = north east ]at(-2*\blockwidth-2*\blockdist,2*\blockheight + 1.5*\leveldist){\Large  $\textbf{Level 1}$};
\node [anchor = north west]at(3*\blockwidth+3*\blockdist,2*\blockheight + 1.5*\leveldist){\Large $\begin{matrix*}[l]\text{$2^2$ lists, each with syndromes of vectors}\\
\text{with length $n/4$ and Hamming weight $n/4$}\\\\
\text{Merged on $\ell = \frac{\nu n'}{\log_2(q)}$ coordinates}\\\\
\text{Average list size $L = 2^{\nu n'}$}\end{matrix*}$};

\draw[very thick, dashed, gray] (-2-2*\blockwidth-2*\blockdist,\blockheight + \leveldist/2 + 2*\blockheight + 2*\leveldist) edge (6.2+3*\blockwidth+3*\blockdist,\blockheight + \leveldist/2+ 2*\blockheight + 2*\leveldist);

\node [anchor = north east ]at(-2*\blockwidth-2*\blockdist,3*\blockheight + 2.5*\leveldist){\Large $\textbf{Level 2}$};
\node [anchor = north west]at(3*\blockwidth+3*\blockdist,3*\blockheight + 2.5*\leveldist){\Large $\begin{matrix*}[l]\text{$2$ lists each with syndromes of vectors}\\
\text{with length $n/2$ and Hamming weight $n/2$}\\\\
\text{Merged on another $\ell = \frac{\nu n'}{\log_2(q)}$ coordinates}\\\\
\text{Average list size $L = 2^{\nu n'}$}\end{matrix*}$};

\node [anchor = north east ]at(-2*\blockwidth-2*\blockdist,4*\blockheight + 3.5*\leveldist){$\Large \textbf{Level 3}$};
\node [anchor = north west]at(3*\blockwidth+3*\blockdist,4*\blockheight + 3.5*\leveldist){\Large $\begin{matrix*}[l]\text{Final list: syndromes of vectors}\\
\text{with length $n$ and Hamming weight $n$}\\\\
\text{Merged on the remaining $r'-2\ell$ coordinates}\\\\
\text{Average number of produced solutions = $N'$}\end{matrix*}$};

\end{tikzpicture}
}
    \caption{Example of the algorithm to solve the small instance, for the case of $b = 3$.}
    \label{fig:my_label}
\end{figure}

As in \cite{DebrisAlazard2019}, we consider here the application of a tree merging structure, in which we start from $2^b$ lists and then pairwise merge them until we are left with a final list of solutions.
Each list if populated with $L$ vectors of length $n' 2^{-b}$ and full Hamming weight $n' 2^{-b}$.
Due to lack of space, we do not report here the full details of the algorithm, but it can be easily seen that it is a simple generalization of the technique described in \cite{DebrisAlazard2019}.
Its operating principle is exemplified in Figure \ref{fig:my_label} for the case of $b = 3$.

We consider $L = 2^{\nu n'}$ and, as in \cite{DebrisAlazard2019}, focus on the use of the so-called \textit{amortized lists}: in each merge we desire output lists having the same (average) size of the input ones.
In other words, given that we start from lists of size $L$, we want to always have lists of the same average size.
This is achieved if, in the first $b-1$ levels, one always merges on $\ell = \frac{\nu n'}{\log_2(q)}$ positions.

The list which is produced as the output of the last level contains the solutions to the problem; note that its average size can be estimated as
\begin{align}
N' & \nonumber = 2^{2\nu n' - \left(r'-(b-1)\ell\right)\log_2(q)}\\\nonumber
& = 2^{\nu (b+1)n' - r'\log_2(q)}\\\nonumber
& = 2^{\left((b+1)\nu-(1-R')\log_2(q)\right)n'}\\\nonumber
& = 2^{\left((b+1)\nu-(1-R')\log_2(q)\right)(1-\varphi) n}\\
& = 2^{\rho(b,\nu, \varphi) n},
\end{align}
where $R' = 1-\frac{r'}{n'} = \frac{R}{1-\varphi}$.
When constructing the initial lists, we also have the constraint that it must be $2^{\nu n'} \leq (q-1)^{n' 2^{-b}}$, from which we obtain
\begin{equation}
\nu \leq 2^{-b}\log_2(q-1).    
\end{equation}
Also, consider that each of the initial lists must contain at least one element, so that we also obtain the following constraint
\begin{align}
b &\nonumber \leq \lfloor \log_2(n')\rfloor\\
& = \lfloor \log_2(n)+\log_2(1-\varphi)\rfloor.
\end{align}
Once the small instance has been solved, we use each found candidate for $e_A$ to construct $e_B$.
The probability that a candidate $e_B$ is valid, i.e., that it has full weight $u$, can be estimated as $\left(1-\frac{1}{q}\right)^u
= 2^{\log_2(1-1/q)\varphi n}$, so that the probability that one iteration of the algorithm is successful is

\begin{align}
\mathrm{Pr}[{\sf{success}}] &\nonumber =
1-\left(1-2^{\log_2(1-1/q)\varphi n}\right)^{N'}\\\nonumber
& \approx \min\left\{1\hspace{1mm};\hspace{1mm} N' 2^{\log_2(1-1/q)\varphi n} \right\}\\\nonumber
& = \min\left\{1\hspace{1mm};\hspace{1mm} 2^{\bigg(\left((b+1)\nu-(1-R')\log_2(q)\right)(1-\varphi) + \varphi\log_2(1-1/q)\bigg) n} \right\}\\\nonumber
& = \min\left\{1\hspace{1mm};\hspace{1mm} 2^{\chi(b,\varphi)n} \right\}\\
& = 2^{n\cdot \min\left\{0\hspace{1mm};\hspace{1mm} \chi(b,\varphi)\right\} }.
\end{align}

We neglect the cost of performing PGE (since it is polynomial in $n$), and consider that lists initialization and merging comes with a cost of $L = O\left(2^{\nu n'}\right) = O\left(2^{(1-\varphi) n}\right)$.
Testing each one of the produced $N'$ solutions for the small instance requires $O(N')$ operations.
So, each iteration of the algorithm requires $O\left(2^{\nu (1-\varphi)n}+2^{\rho(b,\nu,\varphi)n}\right) = O\left(2^{n\cdot\max\left\{\nu(1-\varphi)\hspace{1mm};\hspace{1mm}\rho(b,\nu,\varphi)\right\}}\right)$ operations.
Multiplying this quantity by the average number of performed iterations, that is, the reciprocal of the success probability, we obtain the following cost estimate
\begin{align}
T_{SDP} =  \min_{
\begin{smallmatrix}
\varphi\in\mathbb R, \hspace{1mm}0<\varphi < 1-R\\
b\in\mathbb N, \hspace{1mm}b\leq \lfloor \log_2\left((1-\varphi)n\right)\rfloor\\
\nu\in\mathbb R, \hspace{1mm}0<\nu < 2^{-b}\log_2(q-1)
\end{smallmatrix}}\left\{\frac{2^{\nu(1-\varphi)n}+2^{\rho(b,\nu,\varphi)n}}{\mathrm{Pr}[{\sf{success}}]}\right\}.
\end{align}
Taking into account all the above considerations, it is easily seen that the time complexity of the resulting algorithm can be compactly expressed as $2^{\alpha(\varphi, b, \nu) n}$, where the complexity exponent is computed as
\begin{equation}
\min_{\begin{smallmatrix}
\varphi\in\mathbb R, \hspace{1mm}0<\varphi < 1-R\\
b\in\mathbb N, \hspace{1mm}b\leq \lfloor \log_2\left((1-\varphi)n\right)\rfloor\\
\nu\in\mathbb R, \hspace{1mm}0<\nu < 2^{-b}\log_2(q-1)
\end{smallmatrix}}\left\{\max\left\{\nu(1-\varphi)\hspace{1mm};\hspace{1mm}\rho(b,\nu,\varphi)\right\}-\min\left\{0\hspace{1mm};\hspace{1mm}\chi(b,\nu,\varphi)\right\}\right\}.
\end{equation}

\subsection{Decoding One Out of Many}

In the Decoding One Out of Many (DOOM) setting, one has a set of syndromes $\mathcal S = \left\{s^{(0)}, s^{(1)},\cdots,s^{(N-1)}\right\}$ and wants to find a vector $e$ with some weight requirement and so that $H' e^T\in\mathcal S$.
For the low weight SDP, it is well known that general decoding algorithms receive a speed-up of $\sqrt{N}$.
At the best of our knowledge, the DOOM setting has not been studied for the problem of decoding vectors with large weight.
However, it is likely that DOOM is easier that the standard case in which there is only one target syndrome.
In the scheme in \cite{Baldi2013c}, an attacker can construct a set $\mathcal S$ with the following strategy:
\begin{itemize}
\item prepare a set of $N'$ syndromes, computed as $\widetilde s^{(\Theta)} = \mathcal F(m||\Theta)$, where $\Theta$ is a counter and takes values in $\{0,1,\cdots,N'-1\}$;
\item populate $\mathcal S$ with each $\widetilde s^{(\Theta)}$ and all of its $p-1$ QC shifts.
\end{itemize}
By doing this, the attacker end up with a set $\mathcal S$ of size $N = pN'$: clearly, solving DOOM in this case yields a valid syndrome for the message $m$.

In \sysacro{}, differently from \cite{Baldi2013c}, the matrix $Q$ is replaced by a permutation matrix and $\Theta$ is no longer used for rejection sampling, thus it can be fixed as a one-way function of the message to obtain deterministic signatures.
Consequently, at best, an attacker can build $\mathcal S$ by using only the QC shifts of $s = \mathcal F(m)$.
This implies that $N = p$.
To consider the possible speed-ups coming from DOOM, we assume that the gain is the same that one has in the low-weight regime. 
Hence, we assess the cost of a signature forgery attack as
\begin{equation}
T_{DOOM} = \frac{T_{SDP}}{\sqrt{p}} = O\left(2^{\alpha(\varphi, b, \nu)n-0.5\log_2(p)}\right).  
\end{equation}

\section{System design}
In this section we study the system design and the choice of the relevant parameters.
We also provide a preliminary set of parameters for an instance of \sysacro{} able to achieve 128 bits of classical security.
We remark that such a set of parameters does not result from an optimization procedure, and that many degrees of freedom exist for tuning the system parameters, which can be exploited to design other instances of \sysacro{}  possibly achieving even better tradeoffs and higher levels of security.
This, however, is left for future works.

There are two main factors that affect the design of parameters for \sysacro{}:
\begin{itemize}
    \item The security level, which is based on the difficulty of decoding large-weight vectors, as shown in the previous section.
    \item The average number of attempts needed for generating a valid signature, which depends on the desired density of the signatures (related to the security level) and the distribution $d(x)$ of the entries of $S$.
\end{itemize}

For the latter, in the next section we consider the simplified case in which the entries of $S$ take values in $\left\{0;1\right\}$ with equal probability, that is $d(x) = 0.5 + 0.5x$.
This choice allows an easy modeling of the rejection sampling rate, as shown in the next section.
However, more general choices of $d(x)$ are possible, which may provide stronger security guarantees paid in terms of some slower rejection sampling.
Some examples will be provided in Section \ref{sec:params}, along with the corresponding rejection rates estimated through Monte Carlo simulations.
A more general theoretical analysis of the rejection rate is left for future works.

\subsection{Rejection sampling
\label{sec:rejectionsampling}}

As described in Section \ref{sec:SignatureGen}, the signature generation of \sysacro{} may require some rejection sampling in order to guarantee that each produced signature has maximum weight $n$.
To this end, we need to estimate the probability that a random signature is valid: the reciprocal of this probability corresponds to the average number of attempts needed to generate a valid signature.

To do this, we consider that
$$\sigma = (c+e)S^T = cS^T+eS^T = \widetilde c + \widetilde e,$$
where $c = uG$ with $u$ having weight $m_g$, $G$ is formed by rows with weight $w_g$, $e$ has weight $w$ and $S$ has columns and rows with weight $m_S$.
In particular, $e$ is partitioned as 
$$e = [\underbrace{e'}_{\text{Length $k$}}||\underbrace{e''}_{\text{Length $n-k$}}],$$
where $e'$ is an all-zero vector and $e''$ has weight $w$.
Note that $u$ and $e$ take values in $\{0 ; 1\}$, while $\sigma$ generically takes values in $\mathbb F_q$.
From now on, we consider the simplified case in which also the entries of $G$ and $S$ take values in $\{0 ; 1\}$.
Note that this is not a constraint in \sysacro{}, and is considered in this preliminary work only for easiness of analysis.

We start by deriving the weight distribution of $\widetilde c$.
To this end, we consider that also $c$ has a varying weight: indeed, each of its entries is obtained as the inner product between $u$ and the corresponding column of $G$.
We approximate each column of $G$ as a vector formed by $k$ samples of a Bernoulli variable with parameter $\frac{w_g}{n}$. 
Since $m_g<q$, any entry of $c$ will be equal to $x$ with probability  
\begin{align}
\mathrm{Pr}[c_i = x] & \nonumber = \binom{m_g}{x}\left(\frac{w_g}{n}\right)^x\left(1-\frac{w_g}{n}\right)^{m_g-x}\\
& = f_{\frac{w_g}{n}}\left(m_g,x\right).
\end{align}
When $m_g\ll k$ and $w_g\ll n$, one can approximate the above probability by simply considering that either $c_i = 0$ or $c_i = 1$. 
Consequently, we approximate $c$ as a vector of $n$ samples of a Bernoulli variable with parameter $\rho_c = 1-\left(1-\frac{w_g}{n}\right)^{m_g}\approx \frac{m_gw_g}{n}$.

We now consider the product $\widetilde c = cS^T$: each column of $S^T$ has constant weight $m_S$, so that each entry of $\widetilde c$ is in the form $\sum_{i = 0}^{n-1}c_i s_i$.
Again, we approximate $s_i$ as a Bernoulli variable with parameter $\rho_S = \frac{m_S}{n}$.
Consequently, we have
\begin{align}
\mathrm{Pr}\left[\widetilde c_i = x\right] = \sum_{z = x}^{m_g w_g}f_{\rho_c}(n,z)\sum_{\begin{smallmatrix}x' = x,x+q,\cdots\\
x'\leq z\end{smallmatrix}}f_{\rho_S}(z,x').
\end{align}
We repeat the same reasoning for $\widetilde e = eS^T$, considering that $e$ has constant weight $w$ only in the last $n-k$ entries.
Since $w< q$, we obtain
\begin{align}
\mathrm{Pr}\left[\widetilde e_i = x\right] & \nonumber = \binom{w}{x'}\left(\frac{m_S}{n}\right)^{x'}\left(\frac{1-m_S }{n}\right)^{z-x'}\\
& = f_{\rho_S}(w,x').
\end{align}
We are finally ready to obtain the probability that an entry in $\sigma$ is null, that is
\begin{equation}
\mathrm{Pr}[\sigma_i = 0] = \sum_{x = 0}^{q-1}\mathrm{Pr}[\widetilde c = x]\mathrm{Pr}[\widetilde e = -x].    
\end{equation}
So, the probability to obtain a valid signature is 
\begin{equation}
\mathrm{Pr}[\text{$\sigma$ is valid}] =     \Big(1-\mathrm{Pr}[\sigma_i = 0]\Big)^n.
\end{equation}
The reciprocal of the above probability is the average number of attempts, before a valid signature is generated.

\subsection{Computational complexity}

As we have described in the previous section, the bottleneck in the attacks consists in forgery attacks, whose running time is an exponential function of the code length $n$.
Actually, because of the DOOM setting, the time complexity receives a small speed-up owing to the quasi-cyclicity of the code.
Yet, the speed-up grows with the square root of $p$. 
Also, note that the code length $n$ is linear in $p$.
In practice, we choose $p$ as a very small constant (say, a few hundreds), so that the cost of forgery attacks is essentially linear in $n$.
Then, in practice, to achieve a security level of $\lambda$ bits we need $n = O(\lambda)$.

Key generation requires to compute inversions over $\mathbb F_q$ (since we need to compute the systematic parity-check matrix and $S^{-1}$), which can be performed with cost $O(n^3) = O(\lambda^3)$.
The generation of a signature, instead, requires only\footnote{Note that computing hash functions takes time which does not depend on $n$} vector sums and vector-matrix multiplications, so takes time $O(n^2) = O(\lambda^2)$.
Notice that, in principle, one needs to consider also the average number of rejected signatures before a valid one is obtained.
Yet, when $q$ is large enough, the rejection rate is practically constant (and extremely close to $0$).
This implies that the final rejection sampling has no practical impact on the cost of the signature generation algorithm, which we consequently estimate as $O(\lambda^2)$.
Finally, signature verification clearly comes with the same cost $O(\lambda^2)$ since, again, only hashes and multiplications are involved.


\subsection{System parameters
\label{sec:params}}

Based on the above considerations, let us consider a possible instance of \sysacro{} with the following parameters.

\begin{itemize}
    \item Code length: $n=24000$
    \item Code dimension: $k=12000$
    \item Circulant block size: $p=101$
    \item Field order: $q=127$
    \item Syndrome weight: $w=26$
    \item Weight of the rows of $G$: $w_g=11$
    \item Number of rows of G selected per generated codeword: $m_g=12$
\end{itemize}

For the choice of the distribution of the entries of $\Fq$ in each
row of $S$, we can consider several solutions.
A first simple case is that of considering a uniform binary matrix,
that is, $d(x) = 0.5 + 0.5x$.
Through the analysis reported in Section \ref{sec:rejectionsampling} we can verify that this guarantees a negligible rejection rate for the system parameters 
we are considering.
In fact, for these parameters, the probability that a produced signature is not valid (i.e., does not have full weight) is estimated as $1.44\cdot 10^{-6}$, confirming that the rejection rate can, in practice, be neglected.
For instance, the probability that signature generation has to be repeated for more than two times is lower than $2^{-38}$.

However, we may also consider other distributions for the entries
of $S$ including more elements of $\Fq$.
For example, by considering
$d(x) = 0.5783 + 0.4167 x + 0.0042 x^2 + 0.00083 x^{13}$
we have verified through Monte Carlo simulations that the signature
rejection rate is in the order of $1\%$, which is still largely acceptable.
Including larger values from $\Fq$ in the entries of $S$ inevitably
increases the signature rejection rate.
As another example, we have considered
$d(x) = 0.5775 + 0.4167 x + 0.0042 x^2 + 0.00083 x^{13} + 0.00083 x^{25}$
and verified through Monte Carlo simulations that the signature rejection 
rate in this case is in the order of $98.5\%$.
Considering that the generation of each signature in \sysacro{} is very 
fast, however, such a rejection rate may still be acceptable and result in an overall fast signature generation.

From the above choice of the system parameters, it follows that:
\begin{itemize}
    \item The public key size is $K_s = 2436.6$ kB.
    \item The number of different signatures is $N_s = 2^{263.9}$.
    \item The number of different codewords is $N_c = 2^{133.8}$.
    \item The work factor of attacks based on \ac{ISD} is $2^{132}$.
\end{itemize}

Concerning the PGE+SS solver, we found that the attack is optimized with the following parameters:
\begin{itemize}
    \item $b = 9$;
    \item $\nu = 0.010725$;
    \item $\varphi = 0.493000$.
\end{itemize}
The resulting cost, taking also DOOM into account, is $2^{131.6}$.

\section{Conclusion}
We have introduced \sysacro{}, a code-based digital signature scheme exploiting the hash-and-sign paradigm and the difficulty of performing decoding of large-weight vectors for random-like codes.
The new scheme stems from the previous proposal in \cite{Baldi2013c}, but prevents the applicability of attacks exploiting sparse matrices by replacing them with dense ones. 
This is paid in terms of public key size, which is increased by the use of non-binary codes defined over $\Fq$ instead of the binary ones that were used in \cite{Baldi2013c}.
However, the resulting public keys appear to be in the order of $2.5$ MB for $128$ bits of classical security, which seems a reasonable size.
Moreover, the new system exploits a straightforward signature generation procedure, only based on vector-matrix operations, which makes it intrinsically faster than alternative systems requiring decoding for signature generation.

In this first work we only studied the case of one-time keys, showing that known attacks have exponential complexity and allow devising secure instances under such a setting.
The multiple-time case needs the introduction of an additional rejection sampling stage to tune the distribution of produced signatures, and will be studied in future works.

%
%

\bibliographystyle{splncs04}
\bibliography{bibarchive}

\end{document}